\documentclass[11pt]{article}
\usepackage{moriond,epsfig,graphicx}

\begin{document}
\vspace*{4cm}
\title{HEAVY-FLAVOR PRODUCTION AT RHIC}

\author{Ralf Averbeck}

\address{Department of Physics and Astronomy, Stony Brook University,\\
Stony Brook, New York 11794-3800, USA}

\maketitle\abstracts{Experimental results on heavy-quark production in
proton-proton (p+p), deuteron-gold (d+Au), and gold-gold (Au+Au) collisions
at $\sqrt{s_{NN}} = 200$~GeV at the Relativistic Heavy Ion Collider (RHIC) 
are reviewed.}

Hadrons carrying heavy quarks, {\it i.e.} charm or bottom, are important probes
in high energy hadronic collisions. 
Heavy quark-antiquark pairs are mainly produced in initial hard scattering 
processes of partons.
While some of the produced pairs form bound quarkonia, the vast majority
hadronizes into particles carrying open heavy flavor. 
The latter do not only provide a crucial baseline for quarkonia measurements
but are also of prime interest on their own.
Heavy-flavor measurements in p+p collisions provide an important proving ground
for quantum chromodynamics (QCD).
Because of the large quark masses, charm and bottom production can be treated
perturbatively (pQCD) even at small momenta \cite{mangano93}.
This is in distinct contrast to the production of particles carrying light 
quarks only which can be evaluated within the pQCD framework only for 
sufficiently large momenta.
Systematic studies in p+p and d+Au collisions should be sensitive to the 
nucleon parton distribution functions as well as nuclear modifications of 
these such as shadowing \cite{lin96}.
In Au+Au collisions, heavy quarks present a unique probe for the created hot 
and dense medium.
Important observables in addition to heavy flavor yields are energy loss 
\cite{dokshitzer01,armesto05} and azimuthal anisotropy \cite{lin03,greco04} 
as well as quarkonia suppression \cite{matsui86} or enhancement 
\cite{pbm00,thews01,andronic03}.

At RHIC, the PHENIX and STAR experiments study heavy-quark production 
indirectly via the measurement of electrons from semileptonic decays of 
hadrons carrying charm or bottom.
In addition, STAR directly reconstructs hadronic $D$ meson decays in p+p 
and d+Au collisions. 

\begin{figure}[tbh]
\begin{center}
\includegraphics[width=0.49\textwidth]{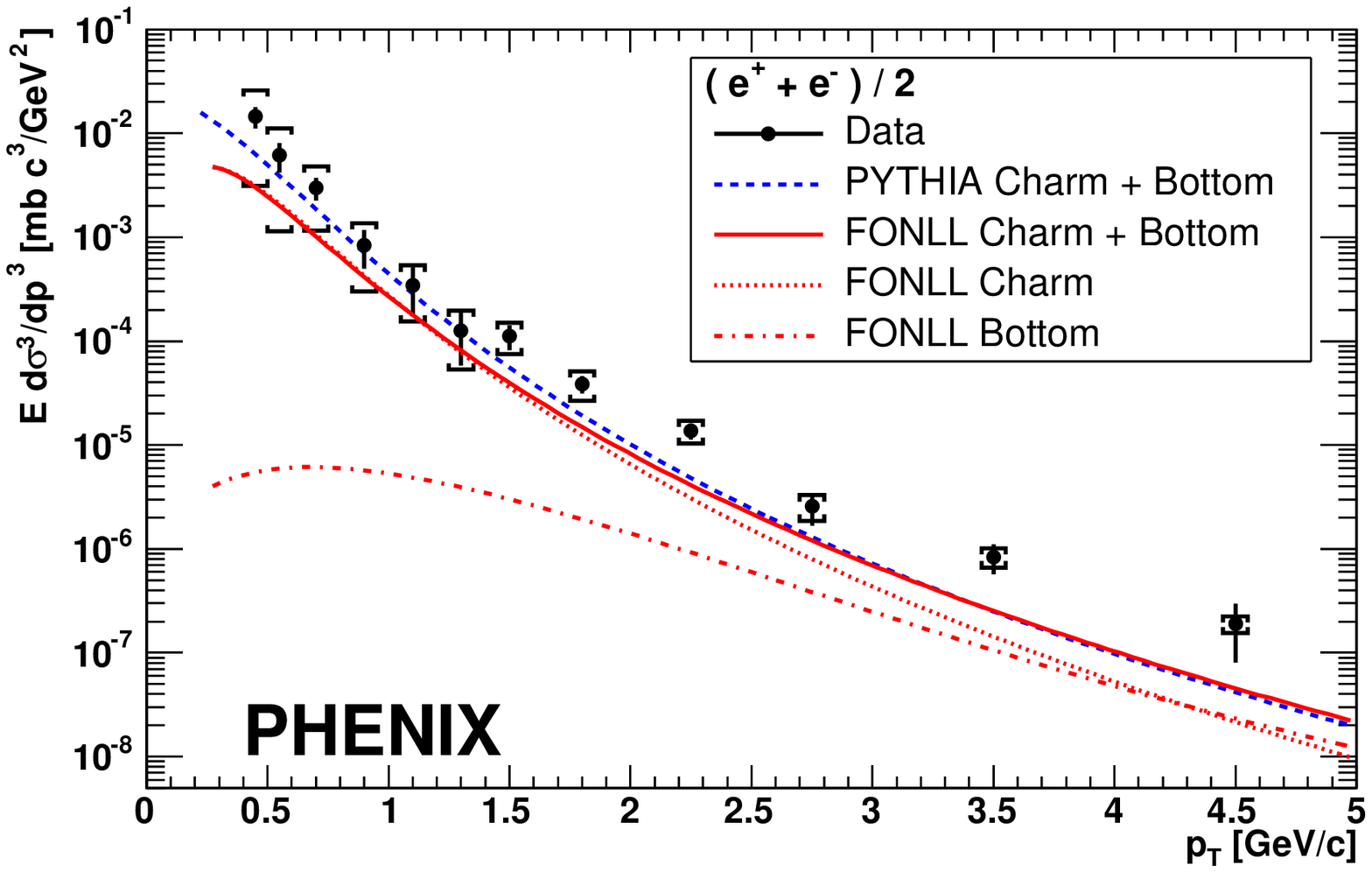}
\includegraphics[width=0.49\textwidth]{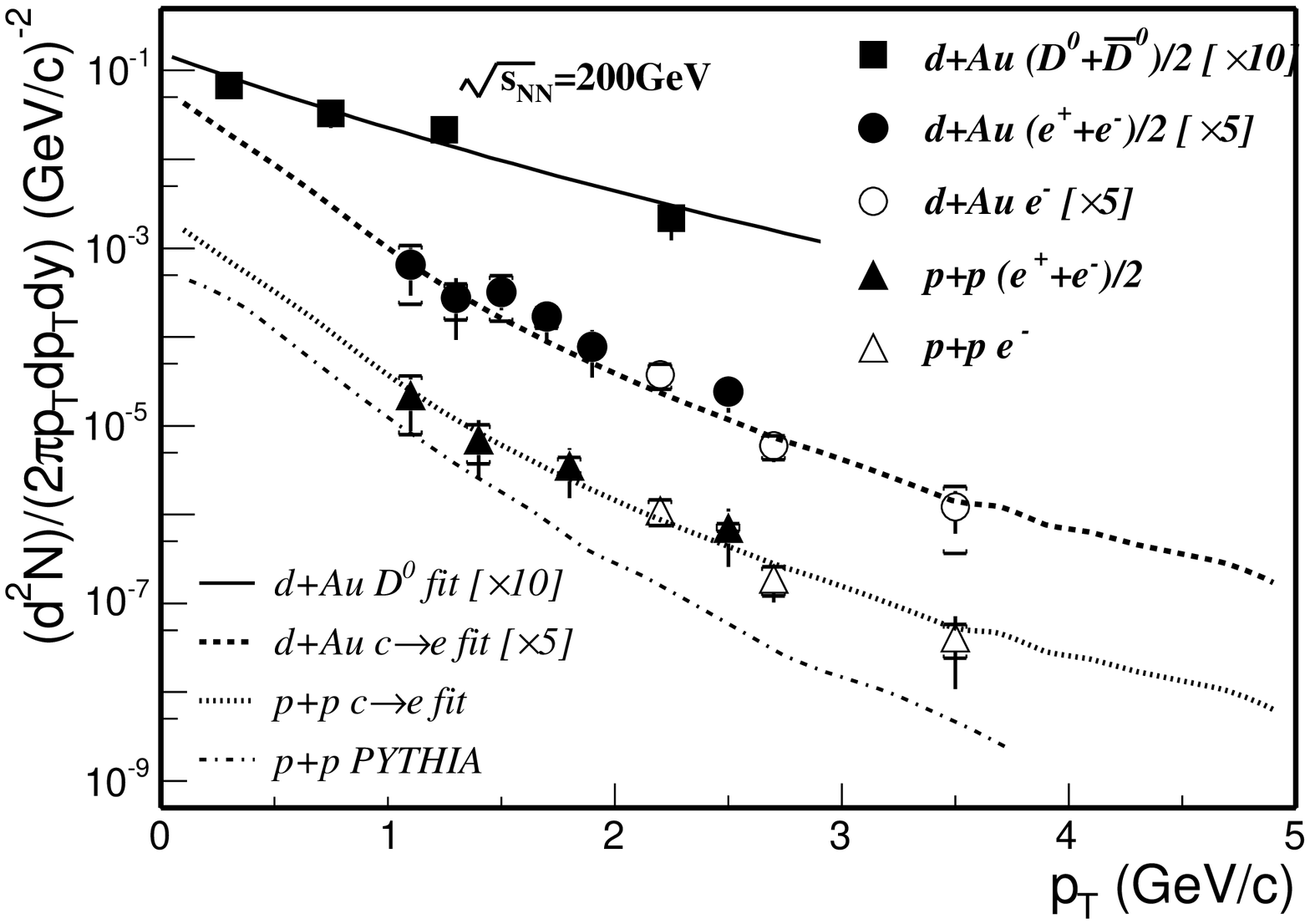}
\caption{Invariant differential cross section of electrons from heavy-flavor 
decays in p+p collisions at 200 GeV in comparision with PYTHIA and FONLL pQCD 
calculations (left panel). Transverse momentum spectra of $D^0$ mesons and 
electrons from heavy-flavor decays in p+p and d+Au collisions at 200 GeV 
(right panel).}
\label{fig1}
\end{center}
\end{figure}

The mid rapidity electron spectrum from heavy-flavor decays measured by PHENIX 
in p+p collisions at $\sqrt{s} = 200$~GeV \cite{phenix_e_pp} is shown in 
Fig.~\ref{fig1} (left panel).
A leading order PYTHIA calculation and a recent next-to-leading order (FONLL) 
pQCD calculation \cite{FONLL} are compared to the data.
The corresponding STAR measurement \cite{star_dau} agrees with these data 
within the substantial uncertainties.
While the spectrum is significantly harder than predicted by PYTHIA, the FONLL
calculation describes the shape better but still leaves room for further
heavy-flavor production beyond the included NLO processes, {\it e.g.} via
jet fragmentation.
Bottom decays are expected to be essentially irrelevant for the electron cross
section at $p_T < 3$~GeV/c and become significant only for $p_T > 4$~GeV/c.

$D^0$ meson spectra, measured by STAR in p+p and d+Au collisions at 
$\sqrt{s_{NN}} = 200$~GeV are shown in Fig.~\ref{fig1} (right panel) together 
with electron spectra from heavy-flavor decays \cite{star_dau}. 
It is important to note that the electron and $D$ meson data are compatible 
with each other, {\it i.e.} the measured electron spectra agree within errors
with spectra calculated for semileptonic $D$ meson decays.
The nuclear modification factor $R_{dA}$ is calculated as the ratio of the
d+Au electron spectrum to the spectrum from p+p collisions scaled with the 
number of underlying nucleon-nucleon binary collisions.
Averaged over the range $1 < p_T < 4$~GeV/c, STAR 
measures\nolinebreak \cite{star_dau} $R_{dA} = 1.3 \pm 0.3 \pm 0.3$, 
which is consistent within errors with binary scaling as expected for a 
point-like hard pQCD process. 
A modest Cronin enhancement is not excluded by the data. 
Electron spectra measured by PHENIX in d+Au collisions as function of 
centrality confirm the observed binary scaling \cite{phenix_e_pp}.

\begin{figure}[tbh]
\begin{center}
\includegraphics[width=0.49\textwidth]{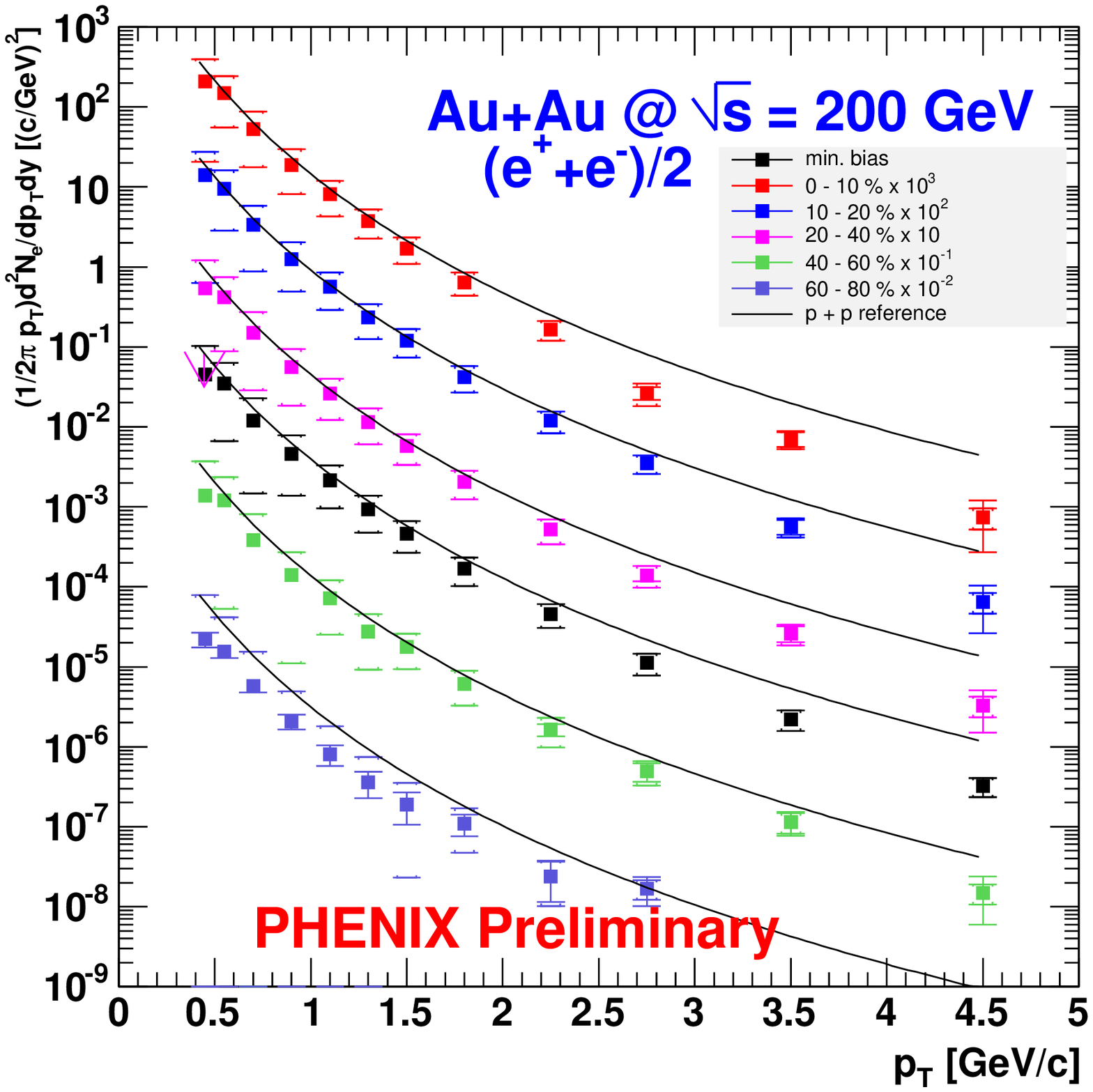}
\includegraphics[width=0.49\textwidth]{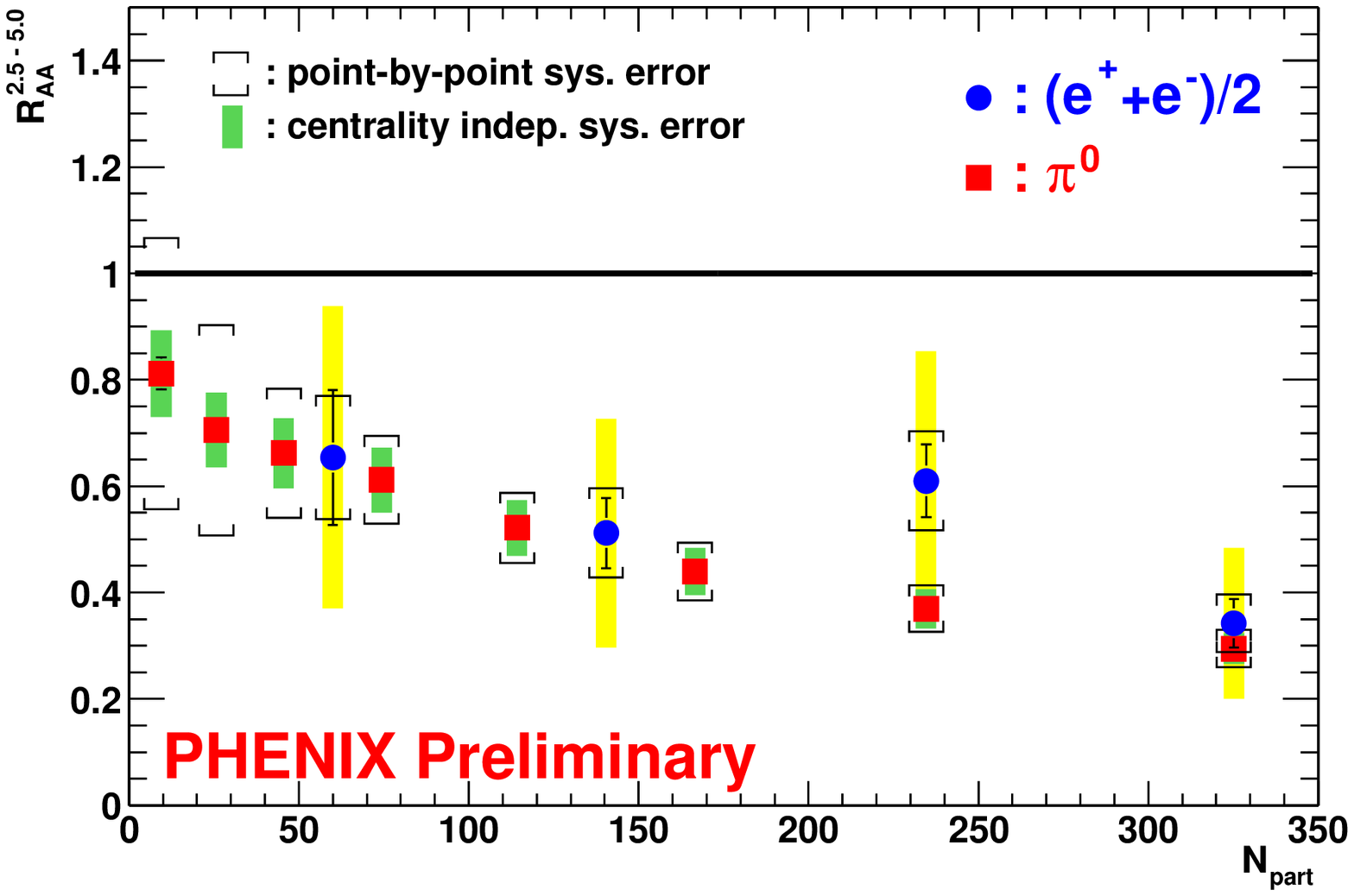}
\caption{Heavy-flavor electron $p_T$ spectra for different Au+Au centrality 
selections at 200 GeV (scaled for clarity) compared to fits to the binary 
scaled p+p measurement (left panel). Nuclear modification factor $R_{AA}$ in 
the range $2.5 < p_T < 5$~GeV/c as function of the number of participant 
nucleons for $\pi^0$ and electrons from heavy-flavor decays (right panel).} 
\label{fig2}
\end{center}
\end{figure}

For Au+Au collisions at $\sqrt{s_{NN}} = 200$~GeV, PHENIX has demonstrated 
\cite{phenix_e_auau} that the yield of electrons from heavy-flavor decays 
is consistent with binary scaling in the range $0.8 < p_T < 4$~GeV/c, which 
is entirely dominated by charm decays.
Preliminary electron data, however, indicate a strong modification of the
spectral shape in central collisions.
Relative to binary scaled p+p data, electrons from heavy-flavor decays are
significantly suppressed at high $p_T$ in central Au+Au collisions as shown in 
Fig.~\ref{fig2} (left panel) \cite{phenix_e_raa}.
This observation is consistent with a scenario where quarks suffer energy
loss while propagating through the hot and dense medium created at RHIC.
The nuclear modification factor $R_{AA}^{2.5-5.0}$, defined as the ratio
of the yield of electrons from heavy-flavor decays in Au+Au collisions in 
the range $2.5 < p_T < 5$~GeV/c to the binary scaled yield in p+p collisions
in the same $p_T$ range, is shown as function of the number of participant 
nucleons $N_{part}$ in Fig.~\ref{fig2} (right panel) together with the same 
quantity for neutral pions as measured by PHENIX \cite{phenix_pion_raa}.
The high $p_T$ electron suppression is comparable to the pion suppression,
but the experimental uncertainties are still substantial and do not allow 
to establish the centrality dependence of heavy quark energy loss.

\begin{figure}
\begin{center}
\includegraphics[width=0.49\textwidth,viewport=0 2 540 398,clip]{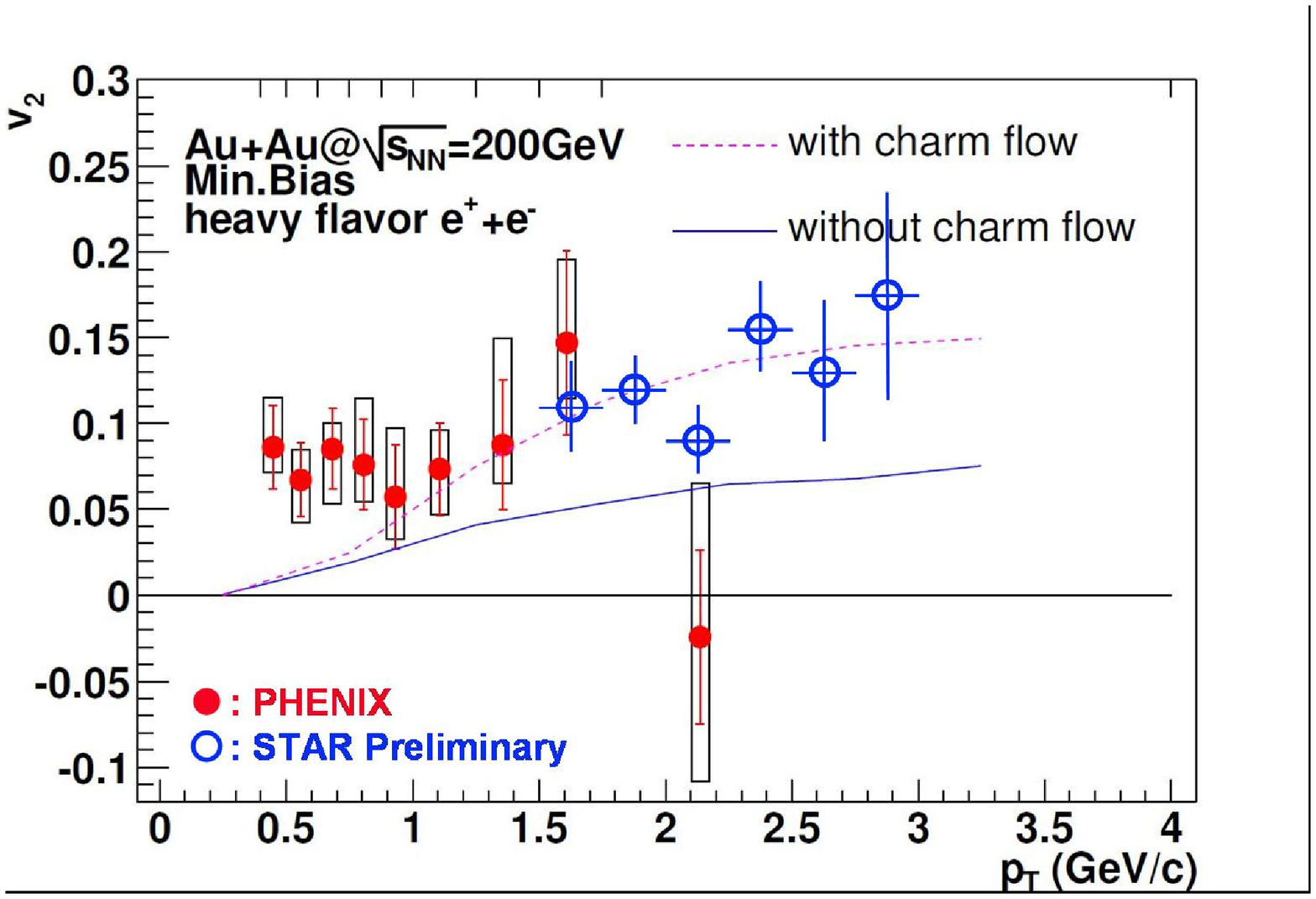}
\caption{Elliptic flow strength $v_2$ of electrons from heavy-flavor decays
in Au+Au collisions at 200 GeV as function of $p_T$ in comparision with 
recombination model calculations with and without charm quark flow.}
\label{fig3}
\end{center}
\end{figure}

A complementary observable related to the interaction of heavy quarks with
the medium created in Au+Au collisions is elliptic flow.
The elliptic flow strength $v_2$ for electrons from heavy-flavor decays
is shown as function of $p_T$ in Fig.~\ref{fig3}.
The PHENIX data \cite{phenix_e_v2} and preliminary results from STAR 
\cite{star_e_v2} are compared with two recombination model calculations 
\cite{greco04}. 
Charm quark flow is consistent with the experimental data but the 
uncertainties are currently too large to exclude a scenario where charmed 
hadrons acquire a non-zero $v_2$ through being produced via the coalescence 
of a non-flowing charm quark and a flowing light quark.

\begin{figure}
\begin{center}
\includegraphics[width=0.49\textwidth]{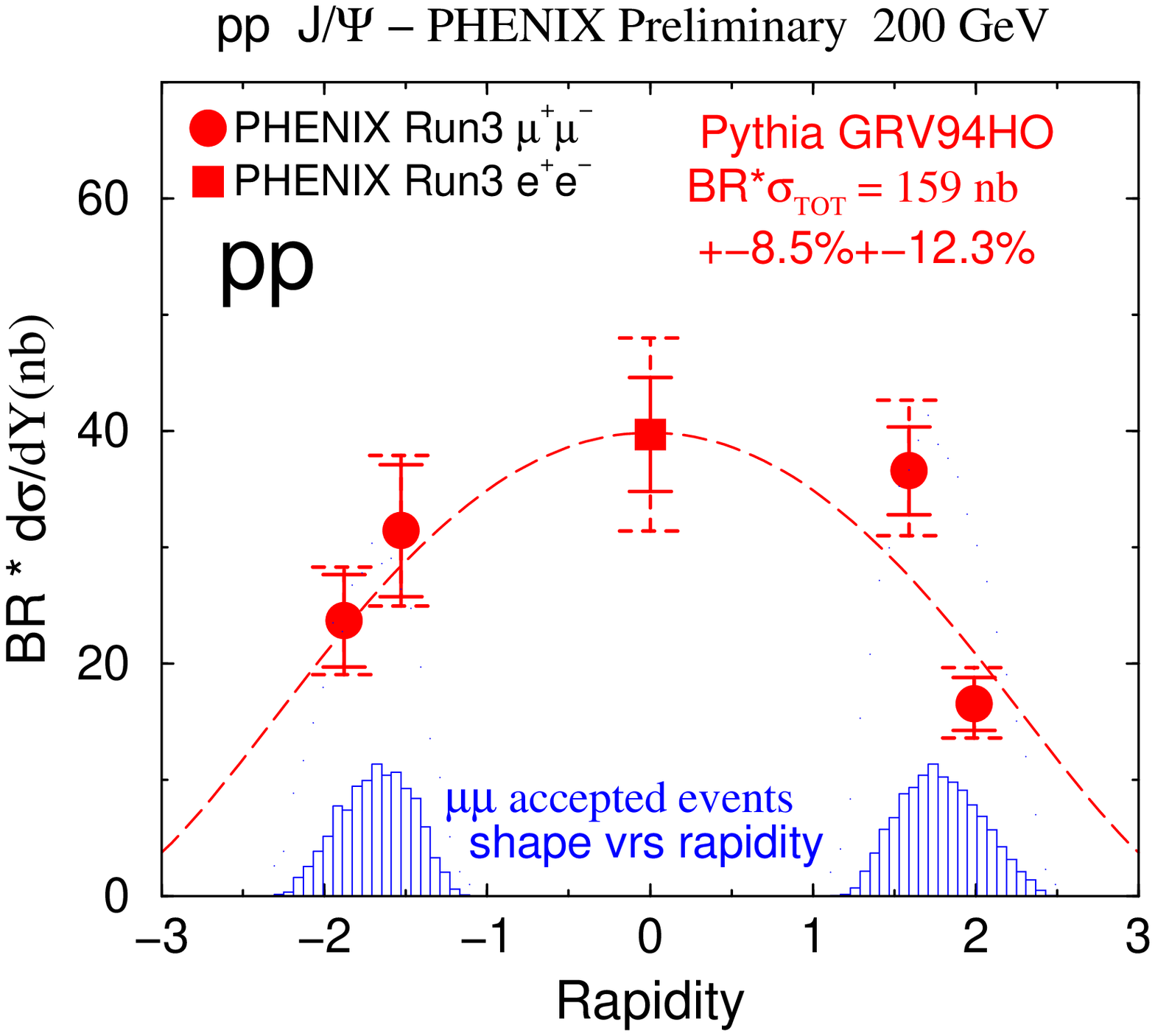}
\includegraphics[width=0.49\textwidth]{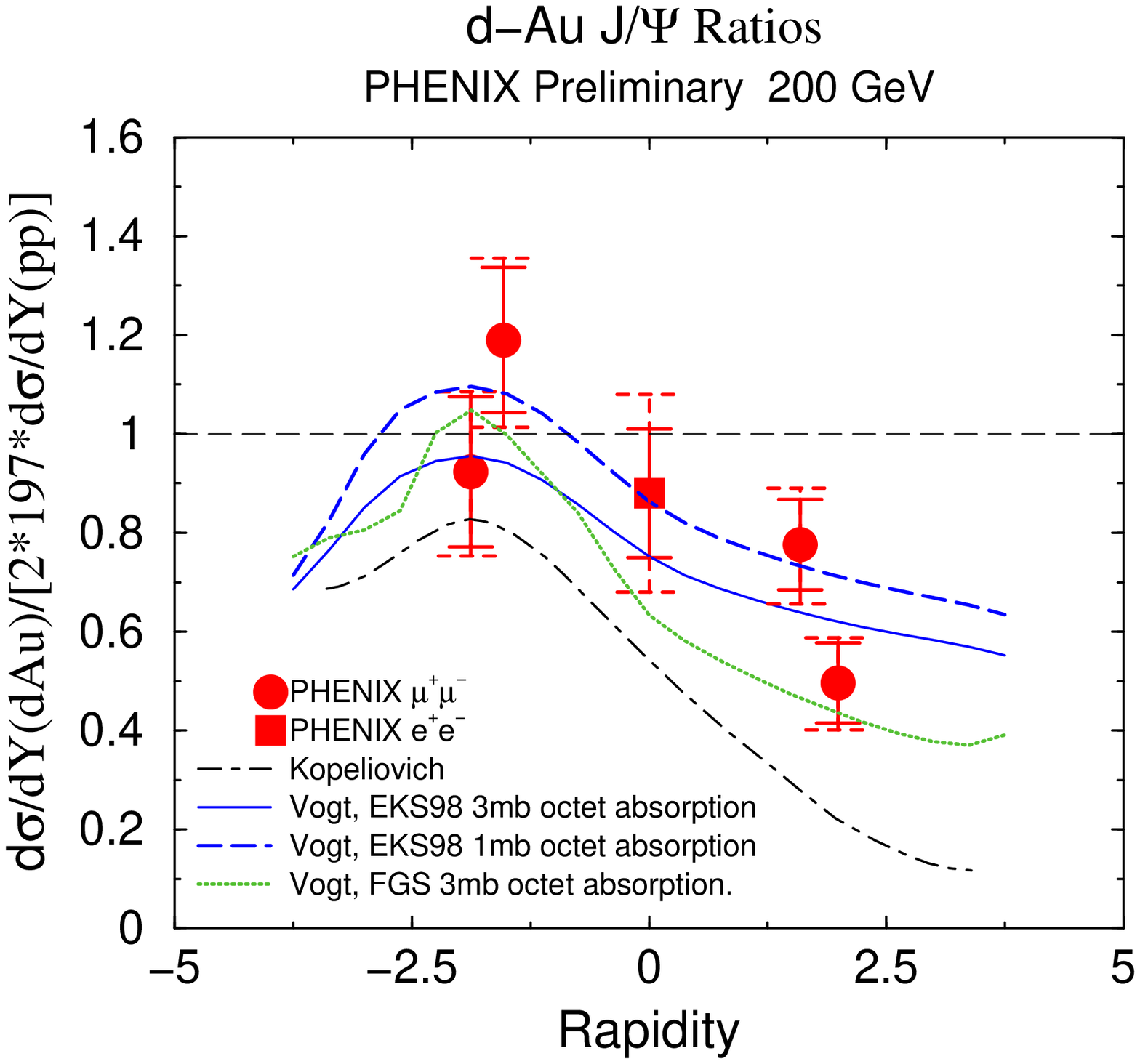}
\caption{J/$\psi$ differential cross section, multiplied with the dilepton 
branching ratio, as function of rapidity in p+p collisions at 200 GeV compared
with a PYTHIA calculation (left panel). Ratio of d+Au and appropriately scaled
p+p J/$\psi$ rapidity distributions compared with model calculations 
$^{21,22}$ (right panel).}
\label{fig4}
\end{center}
\end{figure}

PHENIX has measured J/$\psi$ in the dielectron channel at mid rapidity and in 
the dimuon channel at forward and backward rapidities in p+p and d+Au 
collisions at 200 GeV \cite{phenix_jpsi_pp,phenix_jpsi_dau}.
The shape of the rapidity distribution shown in Fig.~\ref{fig4} (left panel) 
for p+p collisions agrees well with Color Octet Model (COM) and PYTHIA 
calculations using different parton distribution functions.
The measured total cross section is consistent with predictions from COM
and Color Evaporation Model (CEM) calculations.
In d+Au collisions weak cold nuclear matter effect are observed.
The ratio of the rapidity distribution from d+Au collisions to the properly
scaled p+p measurement show in Fig.~\ref{fig4} (right panel) indicates both
weak absorption as well as weak shadowing of the gluon distribution function
in nuclear matter \cite{phenix_jpsi_dau}. 
Given the current uncertainties, it is difficult to disentangle these small
effects.

In a deconfined medium the yield of J/$\psi$ might either be reduced due 
to the expected screening of the attractive QCD potential \cite{matsui86} 
or possibly even enhanced via coalescence \cite{thews01} or statistical 
recombination \cite{pbm00,andronic03}.
While scenarios leading to a strong J/$\psi$ enhancement are disfavored by 
first data from Au+Au collisions at 200 GeV \cite{phenix_jpsi_auau}, a 
definite answer will emerge only from the currently ongoing analysis of the 
high statistics Au+Au data sample recorded in RHIC Run-4.

\end{document}